# Coherent Population Trapping resonances on lower atomic levels of Doppler broadened optical lines

Ersoy Şahin[1,2], Gönül Özen[2], Ramiz Hamid[1], Mehmet Çelik[1] and Azad Ch. Izmailov[3]

[1]*TÜBİTAK National Metrology Institute (UME), Gebze, Kocaeli, TURKEY*

[2]*Istanbul Technical University, Faculty of Science and Letters, Engineering Physics Department Maslak, Istanbul, TURKEY*

[3]*Institute of Physics, Azerbaijan National Academy of Sciences, H. Javid av. 33, Baku, Az-1143, AZERBAIJAN*

[*] *Corresponding author: ersoy.sahin@ume.tubitak.gov.tr*

*Tel: +90 262 679 50 00, Fax: +90 262 679 50 01*

## ABSTRACT

We have detected and analysed narrow high-contrast coherent population trapping (CPT) resonances, which are induced in absorption of the weak probe light beam by the counterpropagating two-frequency pumping radiation. Our experimental investigations have been carried out on an example of nonclosed three level $\Lambda$-systems formed by spectral components of the Doppler broadened $D_2$ line of cesium atoms. We have established that CPT resonances in transmission of the probe beam (in the cesium vapor), at definite conditions, may have not only more contrast but also much lesser width in comparison with well- known CPT resonances in transmission of the corresponding two-frequency pumping radiation. Thus CPT resonances, detected by the elaborated method, may be used in atomic frequency standards and sensitive magnetometers (based on the CPT phenomenon) and also in ultahigh resolution spectroscopy of atoms and molecules.

## 1. INTRODUCTION

At the phenomenon of the coherent population trapping (CPT), a multilevel quantum system subject to decay processes is coherently driven into a superposition state immune from the further excitation, in which the system population is trapped. The CPT is the basis of a number of important applications: ultrahigh resolution spectroscopy, atomic clocks, magnetometry, coherent population transfer among quantum states of atoms (or molecules) and also in some others described, for example in reviews [1-3]. In particular, narrow CPT resonances, detected in absorption of a two-frequency laser radiation (and also in the corresponding induced fluorescence spectrum of a gas medium) on three level atomic $\Lambda$-systems, are successfully applied in atomic frequency standards [4] and in high sensitive magnetometers [5, 6]. For these applications, researchers use, mainly, vapors of alkali atoms (in particular Cs or Rb), whose ground quantum term consists of two sublevels of the hyperfine structure [7]. Resonance excitation of atoms on the $\Lambda$-system scheme is realized by means of the two-frequency radiation from given sublevels (Fig.1). Most such $\Lambda$-systems are not closed because of presence of channels of the radiative decay of the excited state │3 > on some Zeeman sublevels of lower levels │1 > and │2 >, which don't interact with the two-frequency radiation [1-3]. Therefore highly narrow CPT resonances, recorded by known methods on the population of the upper level │3 > (Fig.1) in alkali atoms, have a comparatively small contrast on a more wide spectral background of absorption or fluorescence [4].

Some interesting features of CPT resonances in nonclosed atomic Λ-systems also were detected and analyzed [8, 9]. Important experimental study of the CPT phenomenon in the open Λ-type molecular lithium system ($Li_2$) was presented in paper [10], where CPT resonances were detected in the fluorescence spectrum from the optically excited state │3 > (Fig.1). Moreover theoretical papers have been published on possible CPT applications in such open Λ-systems [11-13]. Thus the possibility of the selective photo excitation of atoms (molecules) within the homogeneous width of their resonance optical lines was shown on the basis of the CPT phenomenon [11, 12]. Moreover the new method of the ultra-high resolution spectroscopy was theoretically suggested, which allowed to identify overlapping optical lines in complex molecular spectra even when frequency intervals between centers of given lines are less than their natural line-widths [13].Given important CPT applications are directly based on the use of a trapped atomic (molecular) populations on lower levels of an open Λ-system.

Indeed, narrow high-contrast CPT resonances may appear in dependences of populations of lower long-lived levels │1 > and │2 > of a nonclosed Λ-system (Fig.1) on the frequency difference ($\omega_2 - \omega_1$) of the bichromatic laser pumping. Let us consider interaction of such a system with 2 monochromatic laser fields. Frequencies $\omega_1$ and $\omega_2$ of given fields are close to centers $\Omega_{31}$ and $\Omega_{32}$ of electrodipole transitions │1 > - │3 > and │2 > - │3 > respectively (Fig.1). The population of this nonclosed Λ-system will be exhausted at intensification of the two-frequency laser pumping with the exception of a fraction of atoms, which may remain on lower levels │1 > and │2 > at the following CPT condition [1-3]:

$$|\delta_2 - \delta_1| \leq W, \qquad (1)$$

where $\delta_1=(\omega_1 - \Omega_{31})$ and $\delta_2=(\omega_2 - \Omega_{32})$ are detunings of laser frequencies. The width $W$ of the CPT resonance in Eq.(1) is determined by intensities of laser fields and by relaxation rates of populations and coherence of quantum states and │1 > and │2 >. Under definite conditions, the value $W$ may be much less than homogeneous widths of spectral lines of optical transitions │1 > - │3 > and │2 > - │3 > (Fig.1). Given nontrivial CPT resonances may be detected by means of an additional probe radiation resonant to a quantum transition from any lower level │1 > or │2 >.

Recently we have carried out experimental research of such CPT resonances for nonclosed three level Λ-systems formed by spectral components of the Doppler broadened $D_2$ line of cesium atoms [14]. Given CPT resonances were detected in absorption of the probe monochromatic light beam under action of the counterpropagating two-frequency pumping radiation. However, in that work we used 2 independent diode lasers as monochromatic light source whose radiations were resonant to different optical transitions │1 > - │3 > and │2 > - │3 > (Fig.1). Therefore detected CPT resonances had comparatively large characteristic widths (about 3 - 4 MHz) because of fluctuations of the frequency difference ($\omega_2 - \omega_1$) for these lasers.

In the present work we applied the same method as in our previous paper [14] but with use of only one diode laser, which generates the monochromatic beam with the stabilized frequency $\omega_2$. The second pumping radiation component (with the frequency $\omega_1$) was obtained from this initial beam by the electro-optical modulator. Therefore the frequency difference $(\omega_2 - \omega_1)$ may be smoothly scaned around the microwave interval (9192.6 MHz) between hyperfine sublevels of the Cs ground term (Fig.2). Thus the spectral resolution of our improved setup was risen at least on 1 order of the magnitude in comparison with the previous work [14]. As a result, new important features of nontrivial CPT resonances have been discovered. In particular we have established that, at definite conditions, CPT resonances of the trapped population on lower levels of the nonclosed atomic $\Lambda$-system in transmission of the weak probe beam may have not only more contrast but also were essentially narrower in comparison with well-known CPT resonances in transmission of the corresponding two-frequency pumping radiation (or the light induced fluorescence of the gas medium).

## 2. EXPERIMENTAL METHOD AND SETUP

Our experimental configuration is shown in Fig 3. The laser frequency $\omega_2$ was stabilized on the cesium transition $6S_{1/2}(F=3) - 6P_{3/2}(F'=3)$ (Fig.2) in the reference Cs cell by using the saturation absorption spectroscopy technique. For obtaining the bichromatic puming beam with both frequencies $\omega_1$ and $\omega_2$ the output of the external cavity diode laser (ECDL) was phase modulated at the $^{133}$Cs hyperfine frequency 9192.6 MHz by an electro-optical modulator (EOM). The sideband-to-carrier ratio was adjusted up to 50% by changing the radio frequency power of the frequency synthesizer, which was tuned in the interval 10 MHz (around the resonant value 9192.6 MHz). This pumping beam was collimated to the diameter $D_{pump} = 5$ mm, linearly polarized by the polarizer ($P_2$) and then sent to the magnetically shielded cell, containing the rarefied Cs vapor at the sufficiently low pressure about 0.1 mPa ($3\times10^{10}$ atom/cm$^3$). The monochromatic probe laser beam was linearly polarized by the polarizer ($P_1$) and then was sent to the given Cs cell in the opposite direction. We analyzed cases for two different values of its diameter $D_{probe} = 5$ mm and 1.8 mm. Given two-frequency pumping and monochromatic probe beams were overlapped in the irradiated Cs cell which was 3 cm length and diameter 2.5 cm. This Cs cell was kept at the temperature of 22 C$^0$ and a residual magnetic field inside the cell was less than 10 mGs. Effective natural and Doppler widths of corresponding cesium optical transitions were about 5.3 MHz and 460 MHz respectively [7]. During the experiment, the intensity $I$ of the bichromatic pumping laser beam may be changed. At the same time the intensity of the monochromatic probe beam was kept constant at the sufficiently low value about 0.01 mW/cm$^2$ in order to avoid its visible nonlinear optical effects. To monitor the CPT resonances in the transmission of the probe beam, flip-flop mirrors ($M_4$, $M_5$) were folded so that they did not reflect while the flip-flop mirror ($M_6$) reflected this beam. To record the CPT resonances in the transmission of the bichromatic pumping beam, the flip-flop mirror ($M_6$) was folded so that it did not reflect while the flip-flop mirrors ($M_4$, $M_5$) reflected the given beam. CPT resonances in transmission for both the probe and pumping beams were detected by the same photodetector PD (Fig.3).

The spectral resolution of this installation was risen at least on one order of the magnitude in comparison with our previous paper [14], where we used two independent lasers (with different frequencies $\omega_1$ and $\omega_2$ ) for receipt of necessary pumping and probe beams. Therefore, unlike paper

[14], in the present work we detected CPT resonances with characteristic widths much less than 1 MHz.

## 3. DISCUSSION OF EXPERIMENTAL RESULTS

According to the level scheme of the Doppler broadened Cs $D_2$ line (Fig.2), the monochromatic component of the laser pumping beam with the stabilized frequency $\omega_2$ effectively interacts with 3 various groups of atoms, whose velocity projections (along the wave vector **k** of this beam) are close to following values [15]:

$$V_{32} = \frac{(\omega_2 - \nu_{32})}{|\mathbf{k}|}, \qquad V_{33} = \frac{(\omega_2 - \nu_{33})}{|\mathbf{k}|}, \qquad V_{34} = \frac{(\omega_2 - \nu_{34})}{|\mathbf{k}|}. \qquad (2)$$

Thus CPT resonances in absorption of the two-frequency pumping are formed by 2 different $\Lambda$-systems $6S_{1/2}(F=3) - 6P_{3/2}(F'=3) - 6S_{1/2}(F=4)$ and $6S_{1/2}(F=3) - 6P_{3/2}(F'=4) - 6S_{1/2}(F=4)$ (Fig.2), which correspond to atomic velocity projections $V_{33}$ and $V_{34}$ in Eq. (2). The counterpropagating monochromatic probe beam with the stabilized frequency $\omega_2 = \nu_{33}$ effectively interacts with pumping radiation only through one common group of atoms, whose velocity projections are close to the zero value $V_{33} = 0$ in Eq.(2). Thus the sole $\Lambda$-system $6S_{1/2}(F=3) - 6P_{3/2}(F'=3) - 6S_{1/2}(F=4)$, mainly, contributes to the CPT resonance in absorption of this probe wave (Fig.2). According to selection rules, at absence of an external magnetic field and orientation of the quantization axis along the same linear polarization of pumping and probe laser beams at our experimental conditions (Fig.3), only optical transitions between Zeeman degenerate Cs levels without change of the magnetic quantum number $m$ were induced [15]. Thus we had 6 nonclosed $\Lambda$-systems (corresponding to magnetic numbers $m=\pm 1$, $\pm 2$ and $\pm 3$) for two resonant adjacent optical transitions: $6S_{1/2}(F=3) - 6P_{3/2}(F'=3)$ and $6S_{1/2}(F=4) - 6P_{3/2}(F'=3)$ (Fig.2), where CPT resonances were formed. Optical repumping of the population of this resultant Zeeman degenerate $\Lambda$-system took place on three magnetic sublevels of the ground Cs term $6S_{1/2}(F=4, m=\pm 4)$ and $6S_{1/2}(F=3, m=0)$, which did not interact with incident pump and probe radiations. Taking into account given features, further we may analyze CPT resonances in absorption of the probe beam on the basis of the simple model of the nonclosed $\Lambda$-system (Fig.1), where quantum states $|1>$, $|2>$ and $|3>$ correspond to Cs levels $6S_{1/2}(F=4)$, $6S_{1/2}(F=3)$ and $6P_{3/2}(F'=3)$ (Fig.2).

Fig.4 presents the narrow CPT deep with the center $\delta_1=\delta_2=0$ in transmission of the probe light beam on the resonant transition $|2> - |3>$ for different pumping intensities. This resonance is caused directly by the trapping of an atomic population fraction of the lower level $|2>$ in the nonclosed $\Lambda$-system (Fig.1) because of a negligible population of its excited state $|3>$ at the CPT condition (1). Well-known CPT peaks in the transmission of the corresponding two-frequency laser pumping are shown in Fig.5. For detected CPT resonances (Figs.4, 5), we have calculated contrasts $C$ (in % with respect to the total recorded background) and widths $W$ (on their half-heights) according to definitions of paper [16]. Corresponding dependences of these values $C$ and $W$ on the intensity of the pumping radiation are shown in figure 6. We clearly detected CPT resonances in absorption of the probe beam at the pumping intensity $I > 2.5$ mW/cm$^2$. Therefore their parameters $C$ and $W$ in figure 6 are not indicated in the interval $0 < I < 2.5$ mW/cm$^2$.

Intensification of the optical pumping causes more essential depletion of populations of lower levels $|1>$ and $|2>$ in the nonclosed $\Lambda$-system (Fig.1) at violation of the CPT condition (1). Then the fall occurs of the background for the increasing CPT resonance in absorption of the probe beam. Similar situation takes place also at duration rise of the optical puming at transits of atoms (in our rarefied gas medium) to the central region of the pumping radiation. Therefore the contrast $C$ of the CPT resonance in transmission of the probe beam increases both at growth of the pumping intensity and at narrowing of the probe beam diameter $D_{probe}$ to the central axis of the counterpropagating pumping beam with the fixed diameter $D_{pump} \geq D_{probe}$ (curves 1 and 2 in Fig.6a). At the same time, nonmonotonic change takes place for the contrast $C$ of corresponding known CPT resonances in transmission of the two-frequency pumping radiation (or fluorescence of the optically excited atomic state) at growth of its intensity, that is the value $C$ decreases after some increase (curve 3 in Fig.6a). Thus the contrast of CPT resonances in absorption of the probe beam essentially exceeds the contrast of CPT resonances in absorption of the two-frequency pumping at its intensity $I > 3$ mW/cm$^2$ (Fig.6a).

It is necessary to note that, even a small angle (~ a few degrees) between linear polarizations of pumping and probe laser beams or a weak external magnetic field (~1 Gs) in the Cs cell (Fig.3) lead to essential decrease of contrasts of given recorded CPT resonances. This is caused by difference of the real system of Zeeman degenerate levels (Fig.2) from the considered model of the Λ-system (Fig.1).

We have established earlier, that the CPT resonance in absorption of the two-frequency pumping radiation is a superposition of CPT resonances formed not only on the considered nonclosed Λ-system $6S_{1/2}(F=3) - 6P_{3/2}(F'=3) - 6S_{1/2}(F=4)$ but also on the additional Λ-system $6S_{1/2}(F=3) - 6P_{3/2}(F'=4) - 6S_{1/2}(F=4)$ (Fig.2). Therefore in our case of the Doppler broadened Cs D$_2$ line, the total width of the given resonance is essentially more that the width of the corresponding resonance in absorption of the counterpropagating probe beam (Fig.6b). Growth of the pumping intensity leads to broadening of these CPT resonances (Figs.4, 5) on quasi-linear dependences shown in figure 6b.

## 4. CONCLUSIONS

On the basis of the elaborated method and corresponding improved experimental setup with the sufficiently high spectral resolution, we have detected and analyzed narrow high-contrast CPT resonances characteristic for a nonclosed atomic (molecular) Λ-systems. These CPT resonances are determined, mainly, by the trapped atomic population on the definite lower level of the Λ-system, from which the resonant optical transition is induced by the weak probe light beam counterpropagating with respect to the two-frequency laser pumping beam. The applied method allows to analyze displays of given nontrivial CPT resonances directly for the definite Λ-system between 3 selected hyperfine sublevels , for example, in the structure of the Doppler broadened Cs D$_2$ line (Fig.2). We have shown, that CPT resonances in the transmission of the probe beam may have not only more contrast but also much lesser width in comparison with known CPT resonances in transmission of the corresponding two-frequency pumping radiation (or fluorescence of the optically excited quantum state). These advantages of given CPT resonances are more obvious at increase of the pumping intensity (Fig.6). Thus detected CPT resonances in transmission of the probe beam may be used in atomic frequency standards and sensitive magnetometers based on the CPT phenomenon and also in ultrahigh resolution atomic (molecular) spectroscopy.

# References

[1] E. Arimondo, Progress in Optics, Elsevier, New York, 1996, Vol. 35, 257-354.
[2] S.E. Harris, Physics Today, July Issue, 1997, 36-42.
[3] K. Bergmann, H. Theuer, B.W. Shore, Reviews of Modern Physics 70 (1998) 1003-1025.
[4] J. Vanier, Applied. Physics B 81 (2005) 421-442.
[5] P.D.D. Schwindt, S. Knappe, V. Shah, L. Hollberg, J. Kitching, L. Liew and J. Moreland, Applied Physics Letters 85 (2004) 6409-6411.
[6] J. Belfi, G. Bevilacqua, V. Biancalana, S. Cartaleva, Y. Dancheva and L. Moi, JOSA B 24(2007) 2357-2362.
[7] A.A. Radtsig, B.M. Smirnov, Reference Data on Atoms, Molecules and Ions, Springer, New York, 1985.
[8] F. Renzoni, W. Maichen, L. Windholz and E. Arimondo, Physical Review A 55 (1997) 3710-3718.
[9] F. Renzoni, A. Lindner, and E. Arimondo, Physical Review A 60 (1999) 450-455.
[10] A. Lazoudis, T. Kirova, E.H. Ahmed, L. Li, J. Qi and A.M. Lyyra, Physical Review A 82 (2010) 023812.
[11] A.Ch. Izmailov, M. Mahmoudi and H. Tajalli, Opt. Commun. 176 (2000) 137-148.
[12] A.Ch. Izmailov, Laser Phys. 15 (2005) 1543-1549.
[13] A.Ch. Izmailov, Laser Phys. 18 (2008) 855-860.
[14] E. Sahin, R. Hamid, C. Birlikseven, G. Ozen and A.Ch. Izmailov, Laser Physics 22 (2012) 1038-1042.
[15] W. Demtroder, Laser Spectroscopy: Basic Concepts and Instrumentation, Springer, Berlin, 2003.
[16] J. Vanier, M. W. Levine, D. Janssen and M. J. Delaney, IEEE Transactions on Instrumentation and Measurement 52 (2003) 822-831.

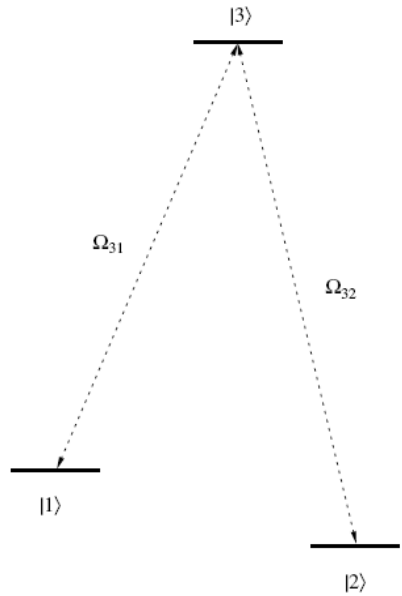

**Fig.1.** The Λ-system of optical transitions $|1>-|3>$ and $|2>-|3>$ (with central frequencies $\Omega_{31}$ and $\Omega_{32}$) between the excited level $|3>$ and long-lived states $|1>$ and $|2>$.

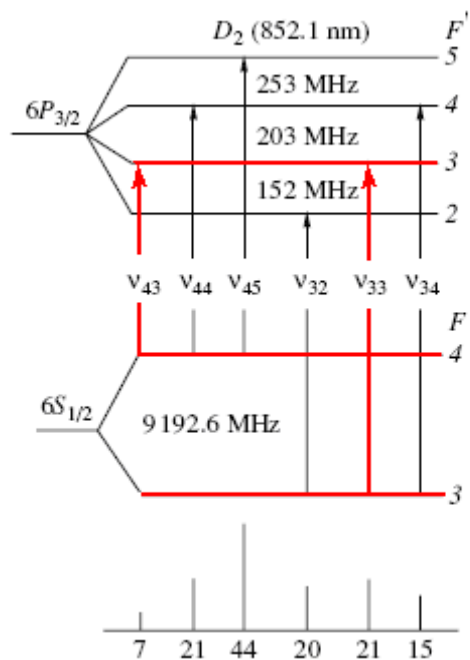

**Fig.2.** Energy level scheme of the $^{133}$Cs $D_2$ line. The relative oscillator strengths of lines, representing hyperfine transitions, are given on the bottom of this figure.

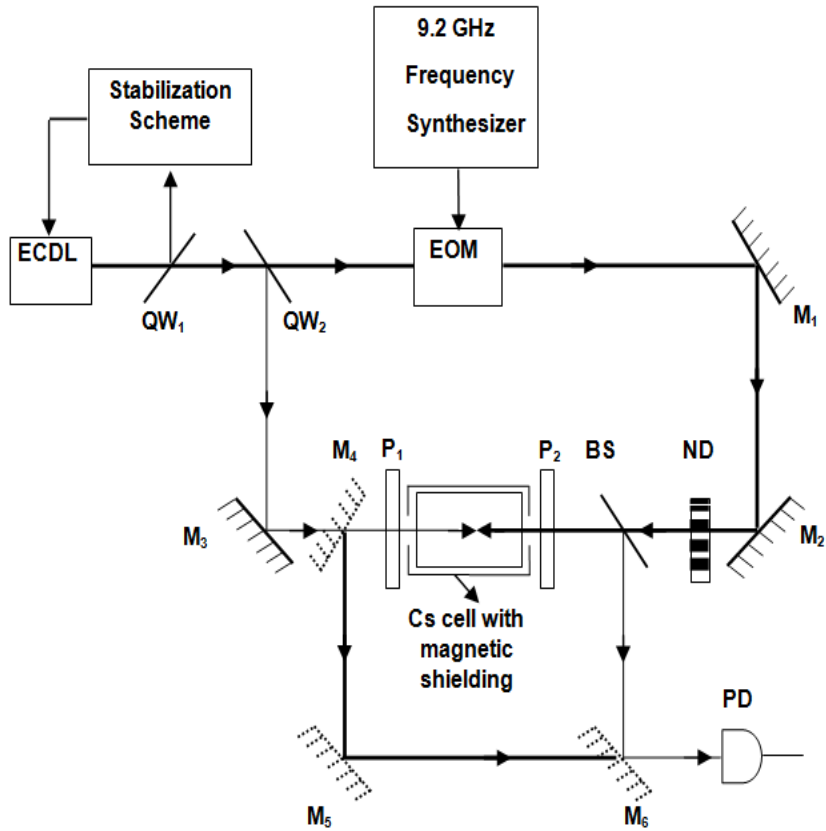

**Fig.3.** Scheme of the experimental setup, which includes external cavity diode laser (ECDL), electro-optical modulator (EOM), beam splitter (BS), mirrors ($M_1$, $M_2$, $M_3$), flip flop mirrors ($M_4$, $M_5$, $M_6$) polarizers ($P_1$, $P_2$), quartz windows ($QW_1$, $QW_2$), neutral density fitler (ND), photodiode (PD) and the Cs cell with the magnetic shielding.

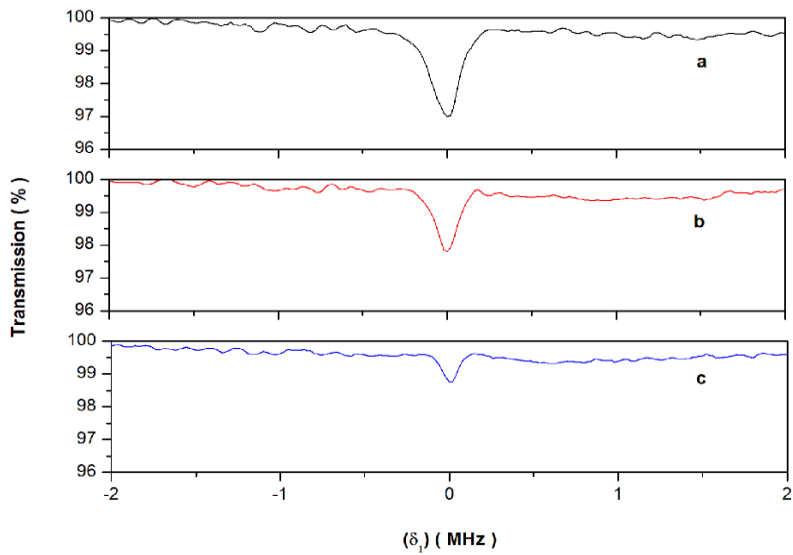

**Fig.4.** Detected CPT resonances in transmission of the probe beam versus the frequency detuning $\delta_1$ at the fixed detuning $\delta_2 = 0$, when the pumping intensity $I = 7.64$ mW/cm$^2$ (a), 5.61 mW/cm$^2$ (b), 3.57 mW/cm$^2$ (c) for equal beams diameters $D_{pump} = D_{probe} = 5$ mm.

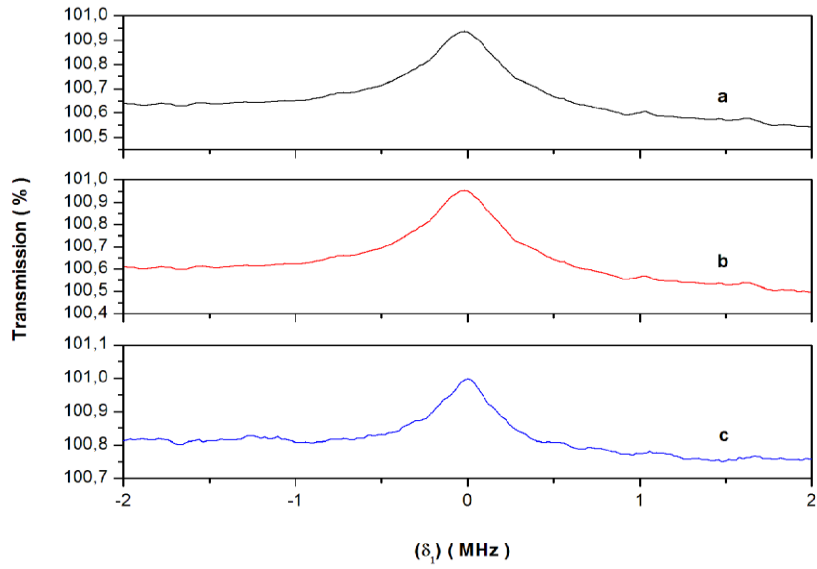

**Fig.5.** Detected CPT resonances in transmission of the bichromatic pumping beam (with the diameter $D_{pump}$ = 5 mm) versus the frequency detuning $\delta_1$ at the fixed detuning $\delta_2 = 0$, when the pumping intensity $I$ = 7.64 mW/cm$^2$ (a), 5.61 mW/cm$^2$ (b), 3.57 mW/cm$^2$ (c).

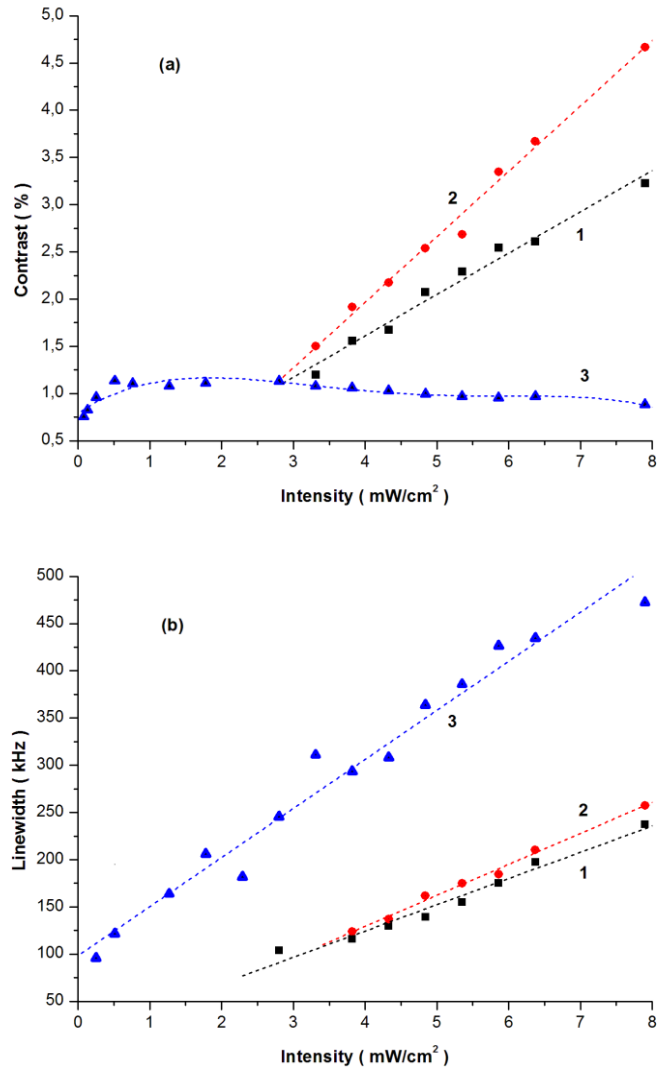

**Fig.6.** The contrast *C* (a) and linewidth *W* (b) of CPT resonances in transmission of the probe beam (curves 1, 2) and the corresponding two-frequency pumping beam (curves 3) versus the pumping intensity *I*, when beams diameters $D_{pump} = D_{probe} = 5$ mm (curves 1) and $D_{pump} = 5$ mm, $D_{probe} = 1.8$ mm (curves 2).